# Comment on "Physics considerations for laser-plasma linear colliders" and on "Beamstrahlung considerations in laser-plasma-accelerator-based linear colliders"

Valeri Lebedev and Sergei Nagaitsev

Fermilab, Batavia, IL 60510, USA

*Abstract*

C. B. Schroeder, E. Esarey, C. Benedetti, and W. P. Leemans {Phys. Rev. ST Accel. Beams **13**, 101301 (2010) and **15**, 051301 (2012)} have proposed a set of parameters for a TeV-scale collider based on plasma wake field accelerator principles. In particular, it is suggested that the luminosities greater than $10^{34}$ cm$^{-2}$s$^{-1}$ are attainable for an electron-positron collider. In this comment we dispute this set of parameters on the basis of first principles. The interactions of accelerating beam with plasma impose fundamental limitations on beam properties and, thus, on attainable luminosity values.

In recent papers [1, 2], Schroeder, Esarey, Benedetti, and Leemans have proposed a set of parameters for a TeV-scale collider based on the plasma wake-field acceleration. To be comparable with conventional designs (such as the ILC and CLIC) the authors chose the design luminosity to be $2 \times 10^{34}$ cm$^{-2}$s$^{-1}$ (Table IV in Ref. [1]). Further, they assumed that the number of particles per bunch is equal to $4 \times 10^9$ and $5.2 \times 10^9$ for plasma densities, $n_0$, of $10^{17}$ cm$^{-3}$ and $2 \times 10^{15}$ cm$^{-3}$, correspondingly. To mitigate the beamstrahlung, they also employed quite short bunches with rms bunch lengths of 1 and 1.3 μm corresponding to the plasma densities mentioned above. These two articles present a fairly complete and detailed analysis of a laser-driven plasma-based collider concept. However, this analysis is of very limited value for several reasons. First, the authors selected the rms beam sizes at the IP to be $\sigma_{IP} = 10$ nm without any justification or discussion of whether it is achievable. Since the luminosity scales as $\sigma_{IP}^{-2}$, it appears that many of the results would be very different had the authors selected a different value of the $\sigma_{IP}$. Second, some of the key collider parameters are missing from these publications. Specifically, beam emittances, beam energy spreads, accelerating synchronous phase, beta-functions at the IP, and beta-functions in plasma are all missing. Without such parameters, the detailed scaling analysis, presented in Ref. [1] and [2], is incomplete and may lead to incorrect conclusions. In this comment we would like to analyze the concept [1, 2] for being self-consistent and to demonstrate that the presented set of collider parameters is unfeasible.

The authors propose to use the quasi-linear regime of acceleration with a low-density beam: "Reduction in the bunch length for fixed charge is limited by bunch density constraints, i.e., $n_b \leq n_0$ to avoid the blow-out regime and the resulting strong beam self-focusing and emittance growth." Let us first examine this statement. In order for the beam density to be lower or equal to the plasma density, the beta-function along the plasma acceleration channel must be equal to or greater than the following value (for a Gaussian distribution):

$$\beta(\gamma) = \frac{N_b \gamma}{(2\pi)^{3/2} \varepsilon_n \sigma_s n_0}, \tag{1}$$

where $N_b$ is the number of particles per bunch, $\gamma$ is the Lorentz factor, $\varepsilon_n$ is the rms normalized beam emittance, and $\sigma_s$ is the rms bunch length. With this beta-function value, the transverse rms beam size, $\sigma_r$, is kept constant along the acceleration channel such that $n_b = n_0$. For parameters of Ref. [1, 2], this transverse beam size would be 50 μm for $n_0 = 10^{17}$ cm$^{-3}$ and 350 μm for $n_0 = 2 \times 10^{15}$ cm$^{-3}$. This is the minimum required beam size. Let us recall that the rms laser spot size, $r_L$, selected by the authors [1], is 70 μm for $n_0 = 10^{17}$ cm$^{-3}$. Thus, the rms beam size in plasma is very close to the rms laser spot size and, according to Fig. 1b [1], the concept would have obvious difficulties with the accelerating field dependence on the transverse particle coordinate. It is also not obvious how to make such weak focusing in plasma in view of the fact that the longitudinal and transverse electric fields are not easily controlled independently of each other in a quasi-linear regime where the overall energy efficiency is important.

Let us now turn our attention to the beam emittance. Ref. [1] concludes that "Coulomb scattering is examined and found not to significantly degrade beam quality". The calculations carried out in the Appendix A [1] are correct. They assume strong plasma focusing, which suppresses the emittance growth. However, both articles suggest a quasi-linear regime, thus implying very weak focusing to attain $n_b \leq n_0$, which greatly amplifies the emittance growth. Therefore, we do not quite agree with the authors' conclusion. The emittance growth in fully-ionized plasma due to the multiple Coulomb scattering is given by (in the ultra-relativistic case):

$$\frac{d\varepsilon_n}{d\gamma} = \frac{2\pi Z(Z+1) r_e^2 n_0 \beta(\gamma) \Lambda_c}{\gamma (d\gamma/ds)}, \tag{2}$$

where $r_e$ is the classical electron radius, $Z$ is the plasma ion charge, and $\Lambda_c$ ($\approx 18$) is the Coulomb log. Combining Eqs. (1) and (2) one obtains the following expression for the emittance growth:

$$\frac{d\varepsilon_n}{d\gamma} = \frac{Z(Z+1) r_e^2 N_b \Lambda_c}{\sqrt{2\pi} \varepsilon_n \sigma_s (d\gamma/ds)}. \tag{3}$$

Assuming that the acceleration rate, $d\gamma/ds$, is constant, and integrating the above equation one finally obtains:

$$\varepsilon_f = \sqrt{\varepsilon_i^2 + \frac{2Z(Z+1) r_e^2 N_b \Lambda_c}{\sqrt{2\pi} \sigma_s (d\gamma/ds)} (\gamma_f - \gamma_i)}, \tag{4}$$

where $\varepsilon_i$ and $\varepsilon_f$ are the initial and final normalized rms emittances, and $\gamma_i$ and $\gamma_f$ are corresponding $\gamma$'s. Setting the initial emittance to zero one obtains the minimum emittance in the low-beam-density ($n_b \leq n_0$) regime to be $\varepsilon_f \approx 10$ μm, where we assumed $n_0 = 10^{17}$ cm$^{-3}$, $Z = 1$ and other parameters from Ref. [1, 2]. This is at least 4 orders of magnitude greater than the estimate obtained in Ref. [1], Appendix A. As one can see from Eq. (4) the emittance growth in the quasi-linear regime does not depend directly on the plasma density. It increases with the decrease of an accelerating rate and, consequently, with the decrease of plasma density; so that it would be about a factor of two larger for $n_0 = 2 \times 10^{15}$ cm$^{-3}$. To achieve a 10-nm rms beam size at the IP with such an emittance, the value of the beta-functions at the IP has to be 10 μm – an obvious challenge for the overall collider concept, especially for the energy spread values discussed in the next paragraphs.

As one can see from the previous paragraph, the authors' concept of a quasi-linear regime with a short bunch results in the following bunch parameters (for $n_0 = 10^{17}$ cm$^{-3}$): the rms bunch length, $\sigma_s (=1 \mu m) << k_p^{-1}$, and the rms transverse beam size, $\sigma_r >> k_p^{-1}$, where $k_p^{-1} = c/\omega_p \approx 17$ μm. This is precisely the beam parameters corresponding to essentially a 1-D plasma wave case. Using the Eq. (64) of Ref. [3], we can estimate the maximum number of particles, $N_{max}$, per bunch at the beam loading limit, defined by a complete cancelation of the accelerating field by the wake-field of a short bunch, and, consequently, implying that particles at the bunch tail see no acceleration. Using the parameters of Ref. [1] and [2] (with $E_z/E_0 = 0.3$) and the transverse rms beam size of 50 μm (see above), we obtain $N_{max} = 3.7 \times 10^9$, which happens to be slightly lower than the value the authors use in their concept for the bunch intensity. For the above parameters the value of the beam loading limit is proportional to $\sigma_r^2$; i.e. the number of particles per bunch has to be much lower if one were to decrease the beam size in order to mitigate the problem with the transverse emittance growth. The authors do not present their own values of the beam size and the beam loading limit. Thus, we can only speculate that there might be a problem with the average energy loss exceeding the accelerating rate. Although one can increase the beam loading limit by increasing the transverse bunch size, there is not much flexibility since the beam size increase will increase the transverse emittance growth, which is already well above acceptable level.

Next, we would like to discuss the longitudinal bunch shaping [4], as suggested by the authors, to achieve the high plasma-to-beam (PB) energy transfer efficiency without an increase of the beam energy spread. The authors assume the PB efficiency to be ~40%. Let us consider an idealized 1-D case as described in Ref. [4]. According to [4], the 40% efficiency could be achieved if the ratio of the electric field at the bunch head to the plasma wave amplitude is: $E_{z1}/E_{z0} = \sqrt{0.6} \approx 0.77$. Here we are using the fact that the bunch head, positioned ahead of crest, does not experience any deceleration, while the bunch tail is located at the wave crest, so that the increase of the accelerating field along the bunch is compensated by the increasing decelerating field of the induced wake-field in the plasma. To attain an exact compensation, both the number of particles per bunch, $N_b$, and the bunch longitudinal density has to be specially adjusted (shaped). Such a procedure allows one to accelerate the maximum number of particles for given beam loading. Thus, for a shaped bunch to have the desired effect of reducing the energy spread, the total bunch length should be equal to $k_p^{-1}\operatorname{acos}\left(\sqrt{0.6}\right) \approx 12$ μm (for $n_0 = 10^{17}$ cm$^{-3}$) to achieve the 40% PB efficiency. Consequently, we estimate that the rms bunch length needs to be at least ~ 3 μm – an obvious disagreement with Refs. [1] and [2]. Increasing the bunch length by a factor of three changes the beamstrahlung parameters and, thus, puts the conclusions of Ref. [2] in question. If one attempts to use short bunches ($<< k_p^{-1}$) for the optimal number of particles per bunch, the resulting total energy spread cannot be made smaller than ~20%. For the beam loading limit estimate obtained above, the resulting momentum spread would be well above this value for the bunch intensity and length suggested in Refs. [1] and [2].

In summary, we believe that the collider parameters, presented in Ref. [1, 2], are not self-consistent. We would also like to note that our attempts to correct the above problems by adjusting the parameters while keeping the same overall performance (i.e., beamstrahlung, luminosity and power consumption) were unsuccessful.

**Acknowledgements**

We would like to thank V. Telnov, N. Solyak, and V. Yakovlev for useful discussions. This research is supported by Fermi Research Alliance, LLC for the U. S. Department of Energy under contract No. DE-AC02-07CH11359.